\documentclass[onecollarge,natbib]{svjour2}
\bibpunct{[}{]}{;}{n}{}{,} % to get "[numbered]" references from natbib
\smartqed  % flush right qed marks, e.g. at end of proof
\usepackage{graphicx}
\newcommand{\be}{\begin{eqnarray}}
\newcommand{\ee}{\end{eqnarray}}

\journalname{Few-Body Systems}
\begin{document}

\title{Photon Generalized Parton Distributions%\thanks{Grants or other notes
}


\author{Asmita Mukherjee         \and
        Sreeraj Nair %etc.
}


\institute{Asmita Mukherjee \at
              Department of Physics \\
              Indian Institute of Technology Bombay\\
              Powai, Mumbai 400076, India
%              \email{fauthor@example.com}           %  \\
}


\maketitle

\begin{abstract}
We present a calculation of the generalized parton distributions of the
photon using  overlaps of photon light-front wave functions.
\end{abstract}

\section{Introduction}
\label{intro}
Generalized parton distributions (GPDs) of the proton are objects of interest both
theoretically and in experiments since some time. Among many other aspects 
that make them interesting, these are a possible way
to get information about the orbital angular momentum of the quarks inside
the proton (GPDs)\cite{rev}. These appear in the amplitudes of certain
exclusive processes for example deeply virtual Compton scattering (DVCS), $ep
\rightarrow ep'\gamma$ where there is a finite momentum transfer between
the intial and the final proton and a real photon is observed in the final
state. It has been shown that in the Bjorken regime, the amplitude factorizes 
and can be written in terms of a hard perturbative part and the soft part 
parametrized in terms of the GPDs. Unlike the ordinary parton distributions (pdfs), 
which can be expressed as forward matrix elements of certain operators, GPDs are
off-forward matrix elements of such  operators, thus they are
richer in content than the pdfs. GPDs are being extracted from data on 
DVCS and hard exclusive  meson production processes at DESY, COMPASS and 
JLab experiments. Apart from $x$, which is the momentum fraction of the 
active quark in the proton, GPDs depend on the skewness $\zeta$, which is 
related to the longitudinal momentum transfer and the momentum transfer 
squared $t$ at a
given scale. The second moment of the sum of the GPDs $H(x, \zeta,t)$ and
$E(x, \zeta,t)$ is related to the part of the nucleon spin carried by the
quarks and antiquarks. In the forward limit the GPD $H(x,0,0)$ gives the pdf
$q(x)$. 
In \cite{burkardt}it was shown that  after a  Fourier transform of
the GPDs with respect to the momentum transfer in the transverse direction
$\Delta_\perp$, one gets the distribution of partons in the transverse
position space or impact parameter plane.  
When the longitudinal momentum transfer is zero, this gives the distribution
of partons in the nucleon in the transverse plane. They are called impact 
parameter dependent parton distributions (ipdpdfs) $q(x,b^\perp)$. In fact
they obey certain positivity constraints and can be interpreted 
as probability densities. In some other works, GPDs have
been investigated in three dimensional position space. In \cite{hadron_optics}
 DVCS amplitude was expressed in boost invariant longitudinal position 
space conjugate to the skewness $\zeta$ and a pattern similar to the diffraction 
pattern in optics was observed.

Deeply virtual Compton scattering (DVCS) $\gamma^* \gamma \rightarrow
\gamma \gamma$ on a photon target was considered in \cite{pire} in the
 kinematic region of large virtuality $(Q^2)$ but small squared
momentum transfer $(-t)$ between the initial and final (real) photon. 
The result was interpreted at leading logarithmic order as a factorized form of
the scattering amplitude in terms of a hard handbag diagram and the
generalized parton distributions of the photon. This was shown  at
leading order in $\alpha$ and zeroth order in $\alpha_s$
when the momentum transfer was purely in the longitudinal direction.
The GPDs show logarithmic scale dependence already in parton model, like 
the photon structure functions. They are interesting  as
they can be calculated in perturbation theory and can act as theoretical
tools to understand the basic properties of GPDs like polynomiality and
positivity. Beyond leading logarithmic order one would need to include the
non-pointlike hadronic contribution which will be model dependent.
 It is interesting to access the partonic structure of   
the photon probed in high energy processes, and photon GPDs 
can shed more light on the partonic content of the photon.

\begin{figure}
\centering
%\begin{minipage}[c]{0.9\textwidth}
\includegraphics[width=7cm,height=6cm,clip]{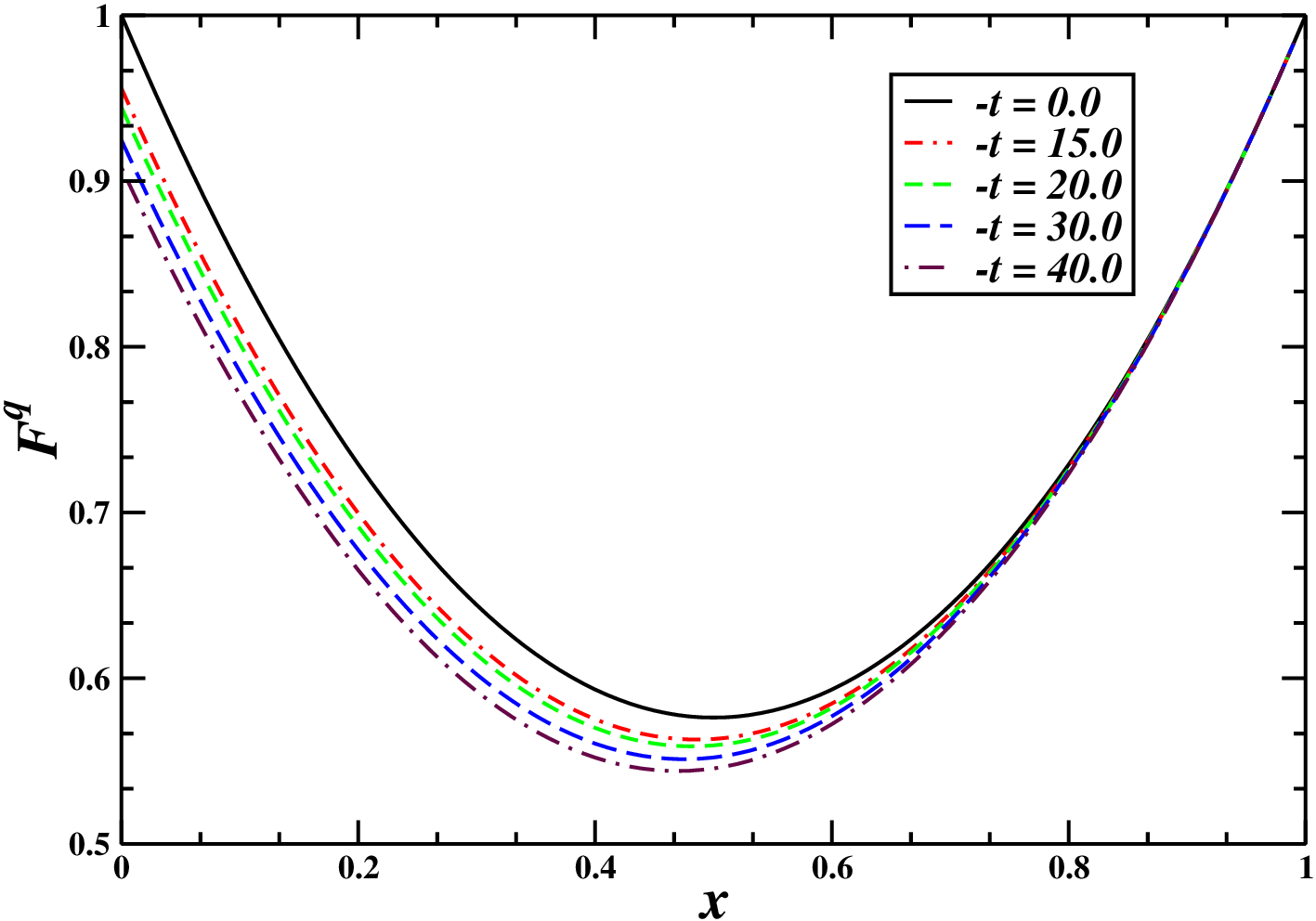}
\hspace{0.3in}
\includegraphics[width=7cm,height=6cm,clip]{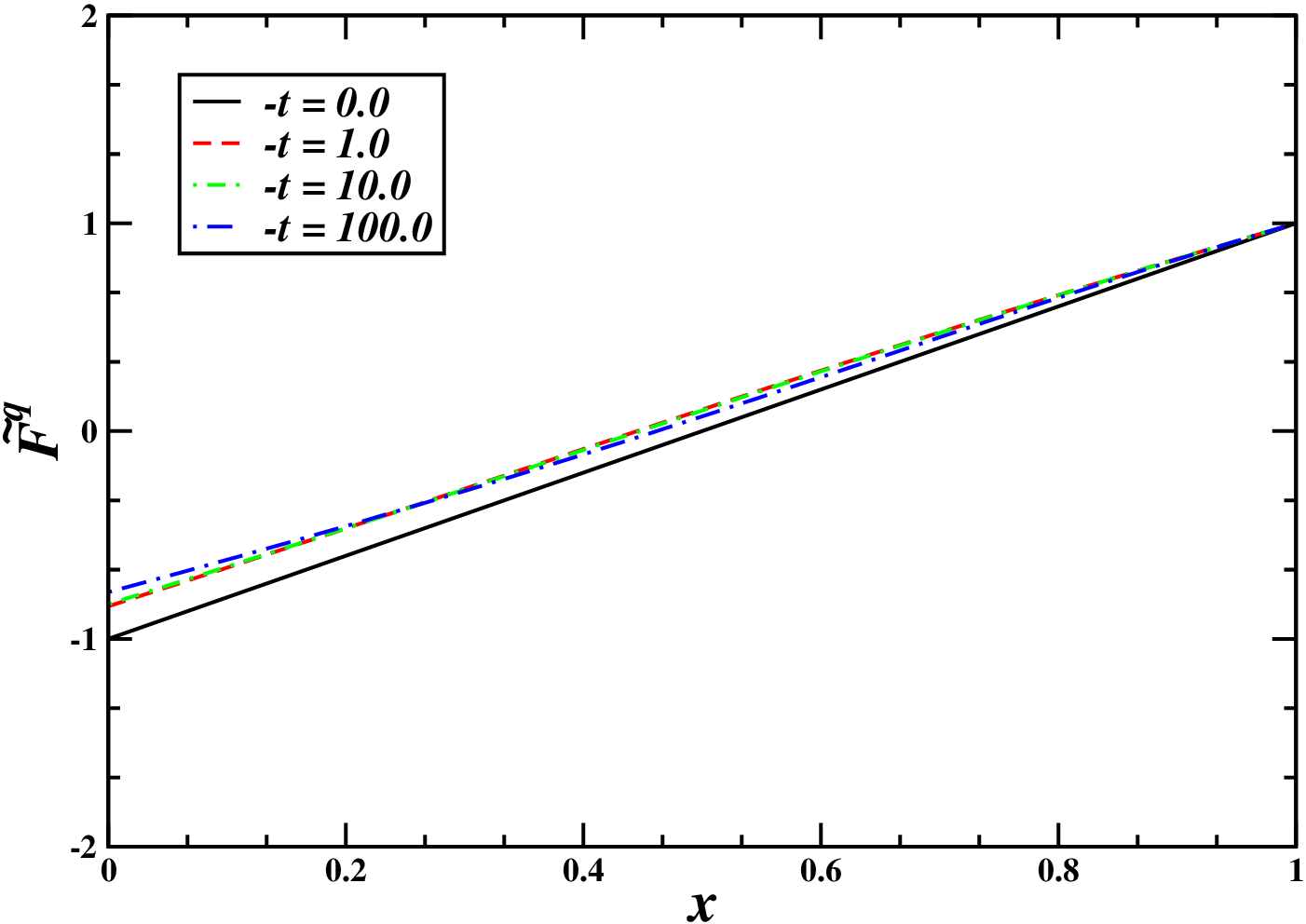}   
%\end{minipage}
\caption{\label{fig1}(Color online) Plots of unpolarized
GPD $F^{q}$ and  polarized GPD  $\tilde{F^{q}}$ vs $x$ for fixed values
 of $-t$ in $MeV^{2}$ $\Lambda = 2\mathrm{GeV}$. In both plots the
normalization factor is chosen to compare with \cite{pire} when $t=0$.}
\end{figure}

\section{Generalized Parton Distributions of the Photon}

The GPDs for the photon are defined as \cite{pire}:
   
\be
F^q=\int {dy^-\over 8 \pi} e^{-i P^+ y^-\over 2} \langle \gamma(P') \mid
{\bar{\psi}} (0) \gamma^+ \psi(y^-) \mid \gamma (P)\rangle ;
\nonumber\\ 
\tilde F^q=\int {dy^-\over 8 \pi} e^{-i P^+ y^-\over 2} \langle \gamma(P')
\mid
{\bar{\psi}} (0) \gamma^+ \gamma^5 \psi(y^-) \mid \gamma (P)\rangle .
\ee
$F^q$ contributes when the photon is unpolarized and $\tilde F^q$ is the
contribution from the polarized photon. The second is extracted from the
terms containing $\epsilon^2_\lambda
\epsilon^{1*}_\lambda-\epsilon^1_\lambda
\epsilon^{2*}_\lambda$ in the amplitude  \cite{pire}. $\lambda$ is the
helicity of the photon.
 We consider the terms where the photon
helicity is not flipped. We have chosen the light-front gauge
$A^+=0$. $F^q$ and $\tilde F^q$ can be calculated using the Fock space 
expansion of the state. Two-particle light-front wave functions of the photon
can be calculated analytically in perturbation theory and are boost
invariant. 

\begin{figure}
\centering
%\begin{minipage}[c]{0.9\textwidth}
\includegraphics[width=7cm,height=6cm,clip]{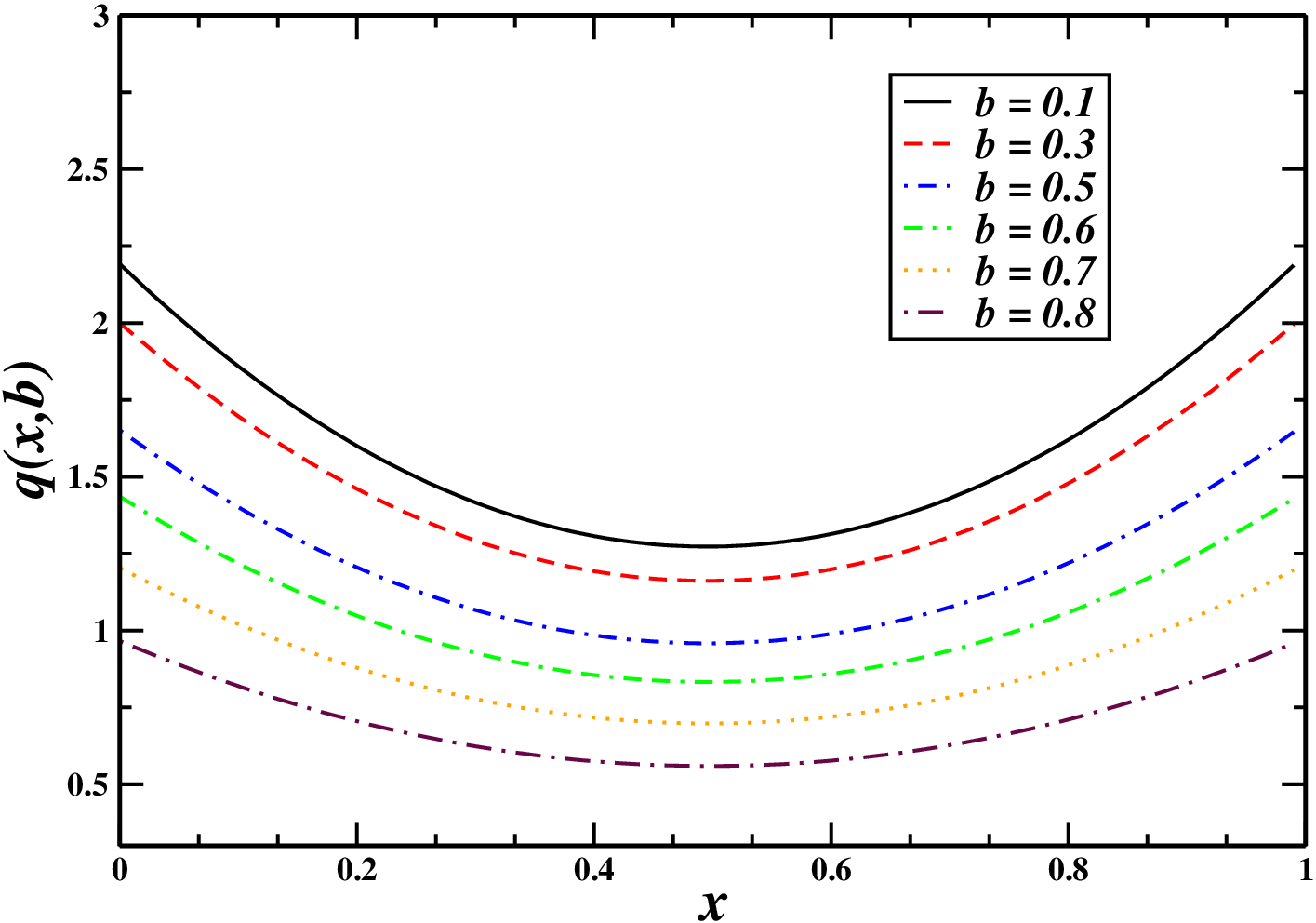}
\hspace{0.3in}
\includegraphics[width=7cm,height=6cm,clip]{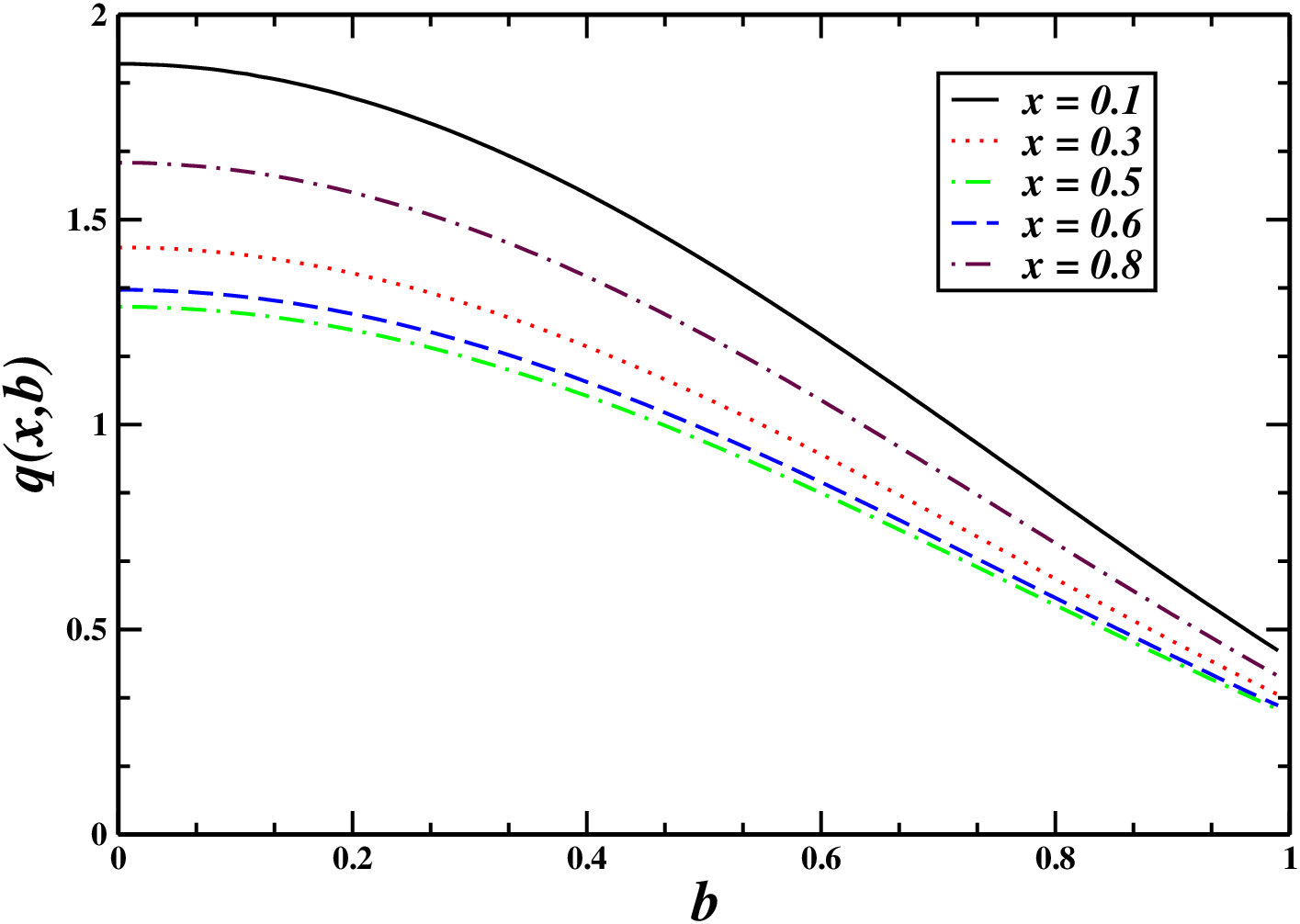}
\caption{\label{fig2}(Color online) Plot of impact parameter dependent  
pdf $q(x,b)$ vs $x$  for fixed $b$ values and  $q(x,b)$ vs $b$ for fixed
values of $x$ where we have taken $\Lambda$ = 2 $\mathrm{GeV}$ and 
$\Delta_{max}$= 3 MeV where $\Delta_{max}$ is the upper limit in
the $\Delta$ integration. $b$ is in ${\mathrm{MeV}}^{-1}$ and  $q(x,b)$
is in ${\mathrm{MeV}}^{2}$ .}
\end{figure}

We calculate the above matrix elements using the 
overlaps of photon light-front wave functions at leading order 
in $\alpha$ and zeroth order in $\alpha_s$, keeping leading 
logarithmic  terms. However, we keep the quark mass terms in the vertex. The
results are scale dependent, this scale dependence in our approach comes
from the upper limit of the transverse momentum integration $ \Lambda = Q$.
There is a lower cutoff on the transverse momentum, which can be taken to  
zero as long as the quark mass is nonzero. Leading order evolution of
the photon GPDs has been calculated in \cite{pire} for non-zero $\zeta$.
The mass terms in the vertex give subdominant contributions. The analytic
expressions for the photon GPDs can be found in \cite{sreeraj}. Here we give the numerical
results.

In analogy with the impact parameter dependent parton distribution of the
proton, we introduce the same for the photon. By 
taking a Fourier transform with respect to the transverse momentum
transfer $\Delta^\perp$ we get the GPDs in the transverse impact  
parameter space.
\be
q (x,b^\perp)&=&{1\over (2\pi)^2}\int d^2 \Delta^\perp
e^{-i\Delta^\perp \cdot b^\perp} F^q \nonumber \\
&=&{1\over 2 \pi}\int \Delta d\Delta J_0(\Delta b) F^q ;
\ee
\be
\tilde q (x,b^\perp)&=&{1\over (2\pi)^2}\int d^2 \Delta^\perp 
e^{-i\Delta^\perp \cdot b^\perp} \tilde F^q \nonumber \\
&=&{1\over 2 \pi}\int \Delta d\Delta J_0(\Delta b) \tilde F^q ;
\ee
where $J_0(z)$ is the Bessel function; $\Delta=|\Delta^\perp|$ and
$b=|b^\perp|$.
In the numerical calculation, we have introduced a maximum limit 
$\Delta_{max}$ of the $\Delta$ integration which we restrict to satisfy
 the kinematics $-t<<Q^2$
\cite{hadron_optics,model}. $q(x,b^\perp)$ gives the
distribution of partonsin the photon. Like the proton, this interpretation
holds in the infinite momentum frame and there is no relativistic correction
to this identification in light-front formalism.

\begin{figure}
\centering
\includegraphics[width=7cm,height=6cm,clip]{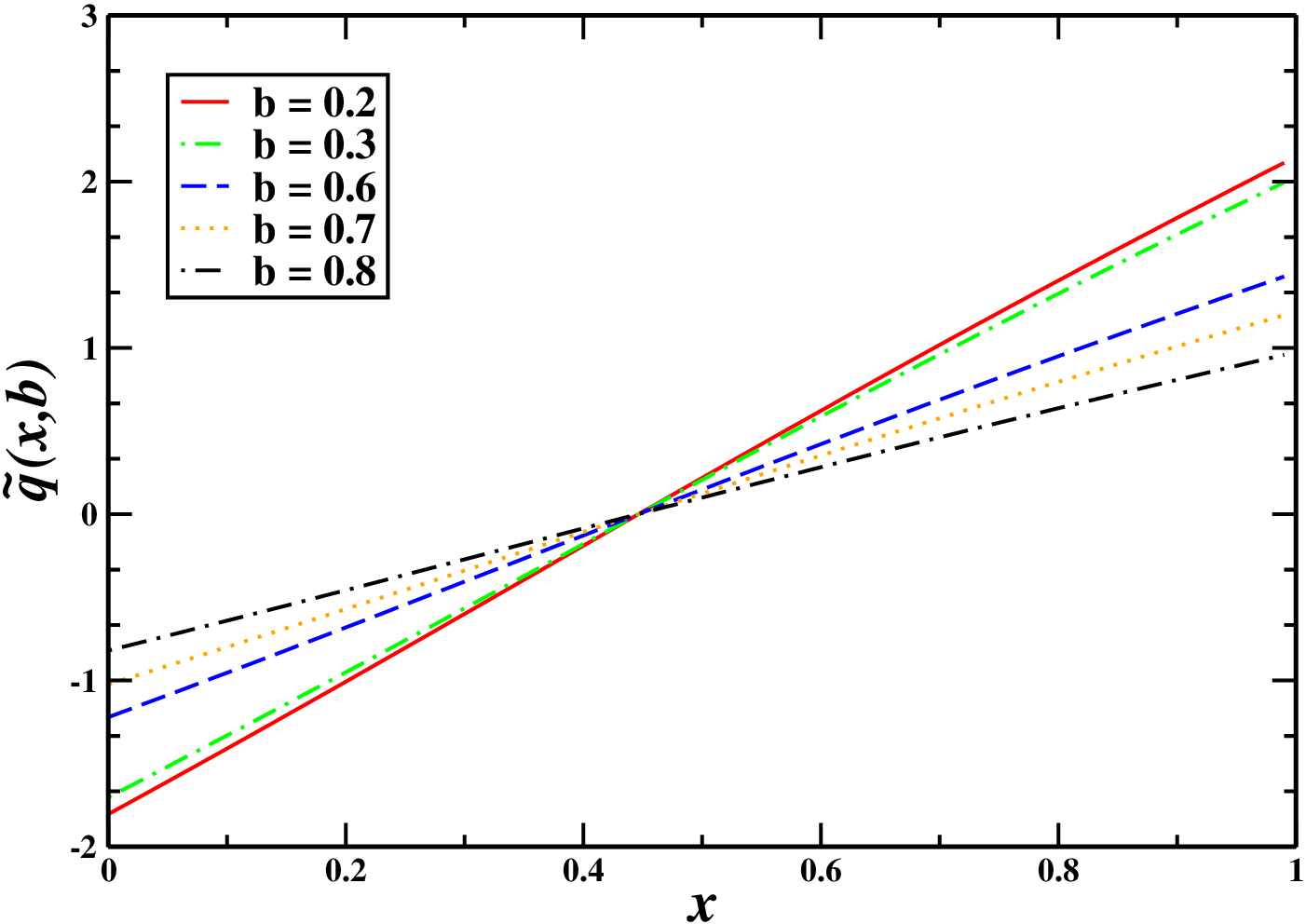}
\hspace{0.3in}
\includegraphics[width=7cm,height=6cm,clip]{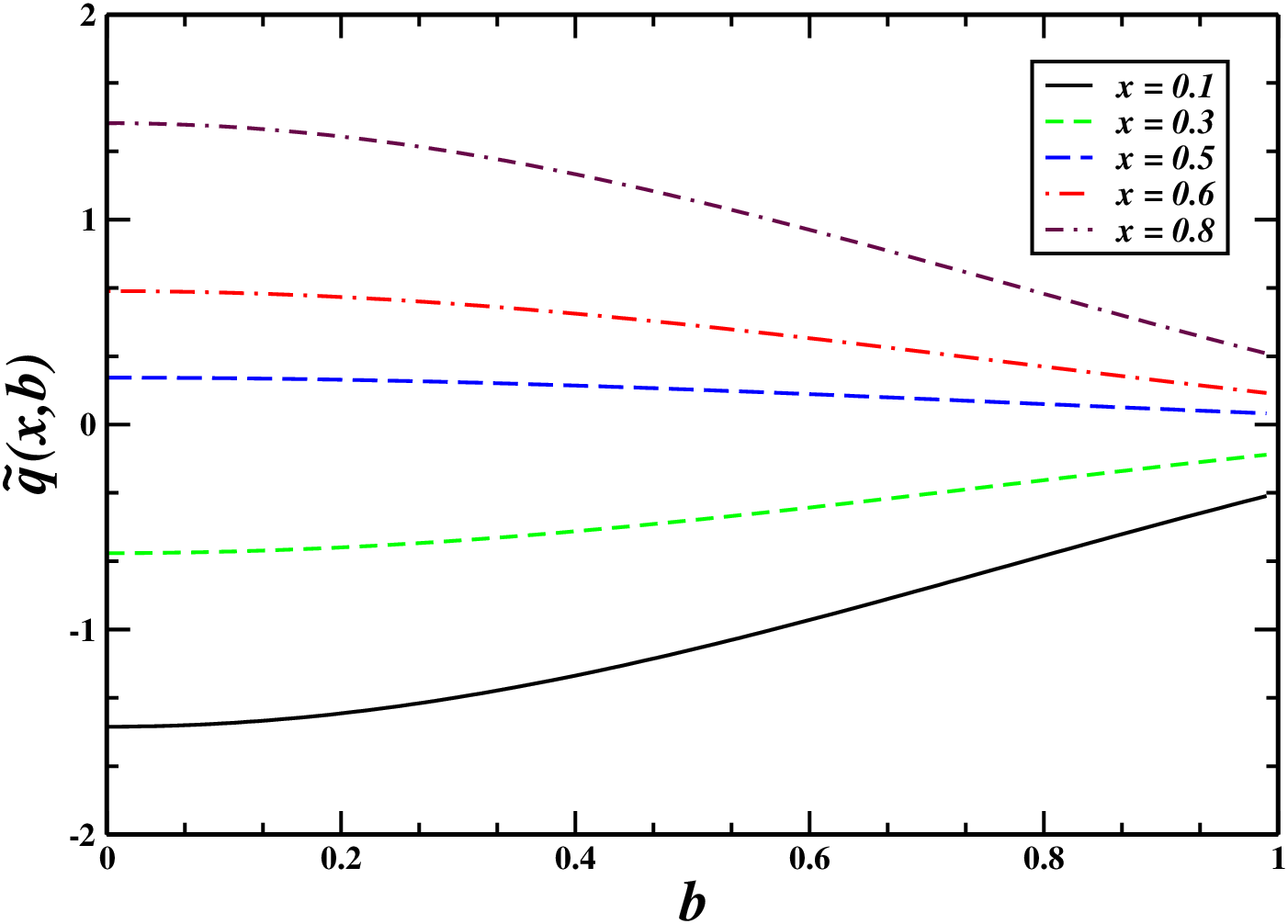}
%\end{minipage}
\caption{\label{fig3}(Color online) Plot of impact parameter dependent
pdfs $\tilde{q}(x,b)$  vs $x$  for fixed $b$ values and $\tilde{q}(x,b)$
vs $b$ for fixed values of $x$
where we have taken $\Lambda$ = 2 $\mathrm{GeV}$ and $\Delta_{max}$= 3
MeV where $\Delta_{max}$ is the upper limit in
the $\Delta$ integration. $b$ is in ${\mathrm{MeV}}^{-1}$ and
$\tilde{q}(x,b)$ is in ${\mathrm{MeV}}^{2}$.}
\end{figure}

We have plotted the unpolarized GPD  $F^q$ and the polarized
GPD $\tilde F^q$ for the photon respectively in Fig. 1 as
functions of $x$ and for different values of $t= -{(\Delta^\perp)}^2$. In
all plots we took the momentum transfer to be purely in the transverse   
direction. We took the mass of the quark as well as antiquark to be $m=3.3$
 MeV;  $\Lambda=Q = 2\mathrm{GeV}$. We
divided the GPDs by the normalization constant to compare with \cite{pire}
in the limit of zero $t$. At larger values of $x$,
most of the momentum is carried by the quark in the photon and the 
GPDs become independent of $t$. The Fourier
transform (FT) of the unpolarized GPD $F^q$ is plotted in Fig. 2 and
polarized GPD $\tilde F^q$ is plotted in Fig. 3. 
In all plots we have  taken $0<x<1$ for which the contribution   
comes from the active quark in the photon ($q {\bar q}$). The smearing in   
$b^\perp$ space reveals the partonic substructure of the photon and shows its     
'shape' in transverse position space. The behavior of the photon GPDs 
in impact parameter space is qualitatively different from phenomenological 
models of proton GPDs
\cite{model}. In the phenomenological parametrization of proton
GPDs where spectator model with Regge-type modification was used, 
the $u$ quark GPDs increase with increasing $x$ for fixed $b$, reaches
a maximum, then decrease. The
peak decreases with increasing $b$ \cite{model}. In the case of a photon
the distribution in $b$ space purely
reveals the internal $q {\bar q}$ structure  of the photon. Near
$ x \approx 1/2$ the peak in $b$ space is very broad which
means that the parton distribution is more dispersed when the $q$ and
${\bar q}$ share almost equal momenta. The slope of the polarized
distribution decreases for higher $b$. 
The sign of the GPD changes at $x=1/2$, at which point the GPD and the pdf
in impact parameter space becomes zero. 

%%%%%%%%%%%%%%%%%%%%%%%%%%%%%%%%%%%%%%%%%%%%%%%%%%%%%%%%%%%%%%%%%%%%%%%%%%%%%%%
\section{Conclusion}
%%%%%%%%%%%%%%%%%%%%%%%%%%%%%%%%%%%%%%%%%%%%%%%%%%%%%%%%%%%%%%%%%%%%%%%%%%%%%%
We discussed a calculation of the generalized parton distributions of
the photon, both polarized and unpolarized. We took the  momentum transfer in
the transverse direction is non-zero. We calculated  at zeroth order in
 $\alpha_s$ and leading order in $\alpha$; also at leading logarithmic
order; we kept the mass terms at the vertex. We took the skewness to be zero.
Taking a Fourier transform
with respect to the momentum transfer in the transverse direction 
we obtain impact parameter dependent parton distribution of the photon. 
These give a unique picture of the photon in transverse position space.
%%%%%%%%%%%%%%%%%%%%%%%%%%%%%%%%%%%%%%%%%%%%%%%%%%%
\section{Acknowledgments}
%%%%%%%%%%%%%%%%%%%%%%%%%%%%%%%%%%%%%%%%%%%%%%%%%%%%%% 
This work is supported by BRNS grant Sanction No. 2007/37/60/BRNS/2913
dated 31.3.08, Govt. of India. AM Thanks the organizers of Light Cone 2011
for the kind invitation and support. 

%%%%%%%%%%%%%%%%%%%%%%%%%%%%%%%%%%%%%%%%%%%%%%%%%%%%%%%%%%%%%%%%%%%%%%%%%%%%

\end{document}